\documentstyle[mnras_cite,epsfig,eqnarray]{mn2e}
\title[Estimates of Black Hole Spin Properties of 55 Sources]{Estimates of Black Hole Spin Properties of 55 Sources}
\author[R. A. Daly]{Ruth. A. Daly \thanks{E-mail:
rdaly@psu.edu}\\
Penn State University, Berks Campus, Reading, PA 19608, USA}
\begin{document}



\maketitle

\label{firstpage}
 
\begin{abstract}
Studies of black hole spin and other parameters 
as a function of redshift 
provide information about the physical state 
and merger and accretion histories of the systems. 
One way that black hole spin may be estimated is through observations 
of extended radio sources.  These sources, powered by outflows from 
an AGN, allow the beam power and total outflow energy to be studied.
In a broad class of models, the beam power of the outflow is related 
to the spin of the black hole. This relationship is used to estimate 
black hole spins for 55 radio sources.  
The samples studied include 7 FRII quasars and 19 FRII radio galaxies with
redshifts between 0.056 and 1.79, and 29 radio sources
associated with CD galaxies with redshifts between 0.0035 and 
0.291.    
The FRII sources studied have estimated spin values of between
about 0.2 and 1; there is a range of values at a given redshift, 
and the values tend to increase with increasing redshift.  Results
obtained for FRII quasars are very similar to those obtained for
FRII galaxies. 
A broader range of spin values are 
obtained for the sample of radio sources associated with CD galaxies studied.
The fraction of the spin energy extracted per 
outflow event is estimated and ranges from about 0.03 to 0.5 for FRII 
sources and 0.002 to about 1 for radio sources associated with CD 
galaxies; the data are consistent with this fraction being independent
of redshift though the uncertainties are large. 
The results obtained are consistent with those predicted by numerical
simulations that track the merger and accretion history of AGN, 
supporting the idea that, for AGN with powerful large-scale outflows, 
beam power is directly related to black hole spin. 

\end{abstract}


\section{INTRODUCTION}

Black hole spin and mass are the fundamental parameters that characterize a 
supermassive black hole.  Black hole spin is related to the merger and 
accretion history of the black hole as discussed, for example, by 
Hughes \& Blandford (2003), Volonteri et al. (2005), 
King \& Pringle (2006, 2007),  Volonteri, Sikora, and Lasota (2007),
King, Pringle, \& Hofmann (2008), and Berti \& 
Volonteri (2008). 
Hughes \& Blandford (2003) consider the coevolution of black hole 
mass and spin in binary merger scenarios and find that the holes are 
spun down by mergers. Volonteri et al. (2005) study the evolution 
of the full distribution of 
black hole spins in hierarchical galaxy formation theories 
including prolonged gas accretion and binary mergers.
Volonteri, Sikora, and Lasota (2007) 
determine and compare spins of black holes in giant elliptical galaxies
with those of disk galaxies and show that elliptical galaxies are 
likely to have higher spins on average than disk galaxies. 
King \& Pringle (2006) suggest that accretion onto supermassive
black holes occurs in a sequence of randomly oriented accretion episodes. 
King \& Pringle (2007) 
show that black hole growth that occurs in a series of small-scale
randomly oriented accretion events leads to black hole spins
that are less than one. 
King, Pringle, \& Hofmann (2008) find that accretion of this type
leads to supermassive black holes with moderate spin, and the 
spin value of an individual black holes may deviate significantly
from the mean value.  In addition, the mean spin value of the 
black hole population 
decreases slowly as black hole mass increases.     
Examples of holes with large spin parameter may occur and 
are most likely to be found in giant elliptical galaxies. 
Berti \& Volonteri (2008) show that the redshift evolution of 
black hole spins for a given population of sources can be used
to determine the merger and accretion history of the sources and whether
the accretion is prolonged or occurs in short-lived chaotic events.  
Thus, studies of spin as a function of redshift may be used as a 
diagnostic of the accretion and merger histories of AGN.   

Spins of individual black holes may be determined by studies quite close to the AGN, or by studies of outflows from the AGN.  Each method of 
measuring black hole spin is  model-dependent.  Spins of three objects have been obtained by studies quite close to the heart of the 
AGN from X-ray measurements. 
Spins obtained using X-ray measurements of AGN are $0.60 \pm 0.07$ for Fairall 9 (Schmoll et al. 2009), $0.6 \pm 0.2$ for AGN SWIFT J2127.4+5654 (Miniutti et al. 2009), and $0.92 - 0.99$ for MCG-6-30-15 (Brenneman \& Reynolds 2006; Reynolds \& Fabian 2008).  
The use of extended radio sources to study black hole spin is discussed and applied to one radio galaxy by McNamara et al. (2009) and to 48 radio galaxies 
by Daly (2009a,b).  Daly (2009a) presents a ``model-independent'' method of placing a lower bound on the black hole spin. The outflow energy is taken as a lower bound on the black hole spin energy, and this bound is combined with the black hole mass to obtain a lower bound on the black hole spin. Powerful radio galaxies were found to have remarkably similar minimum spin values with a weighted mean of $0.12 \pm 0.01$. McNamara et al. (2009) studied the radio galaxy MS0735.6+7421 in the context of the hybrid model of Meier (1999, 2001) and concluded that the outflow is likely powered by the spin energy of a maximally spinning hole.   Daly (2009b) studied 48 extended radio galaxies with redshifts between zero and two, including MS0735.6+7421, in the context of the  Meier (1999, 2001) and Blandford \& Znajek (1977) models of spin energy extraction and obtained an estimate of the spin of each black hole assuming that the magnetic field strength is proportional to the black hole spin; the black hole spins obtained 
range from about 0.1 to 1. The results suggest that the black hole spin 
increases slowly with redshift over the redshift range from zero to two.  

The work presented here on extended radio sources 
expands and improves upon prior studies, which 
included only radio galaxies, in several ways. 
A sample of 7 FRII quasars is included here, and spin estimates 
obtained with these sources are compared with 
results obtained with radio galaxies. 
In the prior study, a magnetic field strength
$B \propto j$ was considered.  Here, two additional characterizations
of the magnetic field are considered, and results obtained with 
different field strength characterizations are compared. 
In addition, the spin energy per unit black hole mass 
is obtained for the 7 FRII quasars, and a new quantity, 
the fraction of the 
spin energy extracted per outflow event is obtained 
and studied for 
7 FRII quasars, 19 FRII galaxies, and 29 radio sources associated with 
CD galaxies.  

The methods of estimating the black hole spin, the 
spin energy per unit black hole mass, and the fraction of spin energy
extracted per outflow event are described in section 2, 
results are presented and discussed in section 3,   
and conclusions follow in section 4.

\section{The Method}

Large-scale radio sources are powered by twin jets that emanate from the AGN.  A class of models have been proposed and developed in which the jets are powered in part or in full by the spin energy associated with a rotating black hole 
and surrounding region
(e.g. Blandford \& Znajek 1977; Rees 1984; 
Begelman, Blandford, \& Rees 1984; Punsly \& Coroniti 1990; 
Blandford 1990; 
Meier 1999, 2001; Koide et al. 2000; McKinney \& Gammie 2004; De Villiers et al.. 2005; Hawley \& Krolik 2006).  In many of these models there is a relationship between the beam power (or energy per unit time) $L_j$ carried by the jet, the black hole mass $M$, spin $j$, and ``braking'' magnetic field strength $B$, which takes the form $L_j \propto B^2M^2j^2$ where $j \equiv a/m$, $a$ is defined in terms of the spin angular momentum $S$, the speed of light $c$, and the black hole mass $M$, $a \equiv S/(Mc)$, and $m \equiv GM/c^2$. For example, this proportionality applies to the Blandford \& Znajek (1977) model (the ``BZ'' model), the modified BZ model discussed by Reynolds, Garofalo, \& Begelman (2006), and the hybrid model of Meier (1999, 2001).

Thus, if the beam power and black hole mass for a given AGN are known, the black hole spin may be studied for various magnetic field strengths (e.g. 
Blandford 1990, Daly 2009b): 
\begin{equation}
j = \kappa (L_{44})^{0.5} ~B_4^{-1} ~M_8^{-1}
\end{equation}
where $L_{44}$ is the beam power in units of $10^{44}$ erg/s, $B_4$ is the poloidal component of the magnetic field that threads the accretion disk and ergosphere in units of $10^4$ G, and $M_8$ is the black hole mass in units of $10^8 \rm{M}_{\odot}$. The constant of proportionality $\kappa$ varies by a factor of a few for different models; for example, in the hybrid model of Meier (1999) $\kappa \approx (1.05)^{-1/2}$, while in the model of Blandford \& Znajek (1977) $\kappa \approx \sqrt{5}$.   

A source with no outflow can still have a significant spin, but only
sources with outflows provide the information necessary to deduce the spin.
For example, in the magnetic switch model of Meier (1999, 2001), the outflow
only occurs when the magnetic field strength reaches a particular value. 

The black hole spin $j$ can be used to determine the black hole 
spin energy $E_s$ in terms of the black hole mass $M$ (e.g. Rees 1984): 
\begin{equation}
E_s/Mc^2= 1-\left({1+[1-j^2]^{1/2} \over 2} \right)^{1/2} ~.
\end{equation}
The fraction $f$ of the spin energy extracted during a particular outflow event may be empirically determined by taking the ratio of the outflow energy $E_*$ to the spin energy: 
\begin{equation}
f \equiv E_*/E_s.
\end{equation}  

Black hole spins, spin energy per unit black hole mass, and the fraction of the spin energy extracted during a particular outflow event are estimated here for sources with empirical determinations of beam power and black hole mass.  To obtain these estimates, three different characterizations of the 
magnetic field strength are considered. For each characterization, the range and redshift evolution of black hole spin and fraction of spin energy extracted are studied. 

The field strengths considered are an Eddington magnetic field 
strength, a constant magnetic field strength, and a magnetic field 
strength that is proportional to the black hole spin. These three
field strengths are related to the black hole properties in 
different ways.  Interestingly, the overall results and general trends 
obtained with each 
characterization are rather similar.  
The Eddington magnetic field strength is 
the field strength such that the energy density of the magnetic field is 
equal to that of a radiation field with an Eddington luminosity (e.g. 
Rees 1984; Dermer, Finke, \& Menon 2008); in units of 
$10^4$ G,  $B_{4,EDD} \simeq 6 M_8^{-1/2}$.  This is the field
strength that is expected for sources radiating at the Eddington 
luminosity.  King (2010) argues that many AGN are likely to be radiating
at this luminosity for much of their lives. To evaluate the impact
of an assumed characterization of the magnetic field strength on 
the quantities studied here, we also consider results obtained assuming
a constant magnetic field strength. This 
possibility has been suggested and considered by several authors 
(e.g. Rees 1984; Punsly \& Coroniti 1990; Blandford 1990) and, more 
recently, Piotrovich et al. (2010).  A characteristic value of 
$10^4 G$, or $B_4= 1$, is generally adopted, and is used here;  
the results obtained here can easily be scaled to any other constant 
value of the field strength.  For comparison, a third characterization of
the magnetic field strength is considered. If it is assumed that 
the fraction of the spin energy extracted per outflow event is constant, 
then observations of FRII radio galaxies indicate that the magnetic
field strength is proportional to the black hole spin,  $B_4 \simeq 2.78 j$ 
(Daly \& Guerra 2002; Daly et al. 2009). Interestingly, this is also
indicated by a comparison of the beam power predicted in the hybrid model
of Meier (1999, 2001) and empirical results obtained by 
Allen et al. (2006) and Merloni \& Heinz (2007).  The results of 
Allen et al. (2006) and Merloni \& Heinz (2007) 
indicate that the beam power is proportional to the 
accretion rate to the power 1.3 and 1.6, respectively, while 
equation (12) of Meier (1999) indicates that the beam power is 
proportional to the accretion rate to the power 1.6 when 
$B \propto j$. Thus, it is possible that the field strength is 
related to the black hole spin, and it is useful to consider this 
case as a comparison.

\section{Results}

\subsection{The Samples}

The method is applied to 7 powerful FRII quasars, 
19 powerful FRII galaxies, and 29 central dominant (CD) galaxies, and includes 
most sources for which the black hole mass and beam power of the outflow have been empirically determined.  The beam powers and black hole masses for the sources are illustrated in Fig. 1 and listed in Tables 1 and 2. The FRII quasars and galaxies have a similar range of beam power and and the quasars have a slightly broader range of black hole mass than the galaxies. Most of the CD galaxies 
have substantially lower beam powers and extend to lower masses than the FRII sources; this is due to the fact that most of the CD sources are quite nearby and can be observed to relatively low flux levels. 

The FRII quasars are obtained from the samples of Leahy, Muxlow, \& Stephens (1989) and Liu, Pooley, \& Riley (1992), and their properties including beam powers are summarized by Wellman, Daly, \& Wan (1997) and Wan, Daly, \& Guerra (2000). All quantities have been computed in a standard spatially flat cosmological model with current mean mass density $\Omega_m = 0.3$, cosmological constant $\Omega_{\Lambda} = 0.7$, and a value of Hubble's constant of $H_0 = 70 \hbox{ km s}^{-1} \hbox{ kpc}^{-1}$. The beam power $L_j$ of the outflow is determined from the pressure of the forward region of the radio lobe, the lobe width, and the rate of growth of the extended lobes by applying the equations of strong shock physics. As noted by O'Dea et al. (2009), this determination of the beam power does not depend upon the offset of the plasma in the radio lobe from minimum energy conditions due to a chance cancellation of the way this offset enters the determination of the beam power.  The total energy that will be expelled by the outflow over its entire lifetime, $E_*$, is obtained from the beam power using the relation established for powerful radio galaxies $E_* \propto L_j^{1/2}$ (O'Dea et al. 2009; Daly et al. 2009; Daly \& Guerra 2002). Quantities are obtained for each side of a source, and the total value is taken to be twice the weighted mean of that for the two sides of the source.  

The sample of 19 FRII radio galaxies is obtained from sources studied by Leahy, Muxlow, \& Stephens (1989), Liu, Pooley, \& Riley (1992), Guerra, Daly, \& Wan (2000), and O'Dea et al. (2009).  Their properties including beam powers are summarized by O'Dea et al. (2009), and are obtained in the same manner as described above for FRII quasars.

The black hole masses of the FRII quasars are obtained from McLure et al. (2006), those for the FRII radio galaxies are obtained from Tadhunter et al. (2003), McLure et al. (2004), and McLure et al. (2006), and those for the CD galaxies are obtained from the values $M_{BH,L_K}$ listed by Rafferty et al.  (2006) after removing the correction factor of 0.35 to bring the black hole masses of Cygnus A and M84 into agreement with those of Tadhunter et al. (2003) and Maciejewski \& Binney (2001).

The CD sample is obtained from Rafferty et al. (2006), and consists 
primarily of FRI sources (Fanaroff \& Riley 1974). 
The total outflow energies $E_*$ are estimated by Rafferty et al. (2006) 
from the pressure and volume of the cavity that is occupied by the radio emitting plasma, and the 
beam powers $L_j$ are obtained by dividing the outflow energy by the buoyancy timescale $t$.  
The authors explain that this timescale is probably an overestimate of the true timescale, 
in which case the beam powers would increase, 
while the energy of the cavity could be due to multiple outflow events, 
which would cause the outflow energies and beam 
powers to decrease.

The source Cygnus A (3C 405) appears in both Table 1 and 2.  
The values listed in each table are obtained using independent methods. 
The beam power of $(47 \pm 8) \times 10^{44}$ erg/s 
listed in Table 1 is obtained by applying the 
equations of strong shock physics to the forward region of the source 
(Wan, Daly, \& Guerra 2000; modified to account for the different cosmological
model adopted), while the value of $(13 \pm 7) \times 10^{44}$ erg/s
listed in Table 2 is obtained by dividing the energy required for the plasma to 
create and occupy 
the lobes by the buoyancy timescale (Rafferty et al. 2006); Rafferty et al. (2006)
explain that the beam power obtained in this way may be an underestimate since the 
buoyancy timescale may be an overestimate. Given that there is only one source
that is common to both methods, it is not possible to do a detailed comparison of
the methods, and in order to compare results obtained with a given method, values
obtained with that method are not modified.  The black hole mass estimates for Cygnus A 
listed in Tables 1 and 2 are also
obtained with two methods, and are in reasonably good agreement. 

The FRII galaxies and quasars studied here are the most powerful FRII radio sources at their respective redshift, and the hosts have been identified as massive elliptical galaxies (e.g. Lilly \& Longair 1984; Best, Longair, \& Rottgering 1998; McLure et al. 2004). Thus, the powerful FRII and CD sources studied are likely to be drawn from the same parent population. 
Since the 
FRII sources represent the most powerful sources are their respective 
redshifts, they likely represent the envelope of the distribution, while the CD sample contains sources at low redshift with a very broad range of radio powers.

The radio emitting regions of the sources studied are quite large
(typically larger than the optical size of the host galaxy), and
the timescale of the outflows are estimated to be millions to 
hundreds of millions of years (e.g. Rafferty et al. 2006; O'Dea et al. 2009).
Thus, it is the very long term average properties of the outflows that
are studied.  
The beam powers are obtained by applying the strong
shock equations for FRII sources, and the pressure confinement 
equations for  
radio sources associated with CD galaxies in clusters of galaxies,
so beam power determinations do not rely on an empirically or theoretically 
determined relationship between radio power and beam power.  
The sources are very large, with 
smoothly varying radio emission; the 
synchrotron radiation is not Doppler boosted or beamed, 
so there are no relativistic correction factors of this type.
In addition, the overall rate of growth of the sources is not
relativistic (e.g. O'Dea et al 2009; Rafferty et al. 2006).   
Thus, the use of the large extended radio emitting regions around
AGN to determine the beam power is relatively simple and straight forward. 
In this sense, the AGN outflows studied here are quite different
from those associated with X-ray binaries.  Outflows from X-ray binaries
can be highly variable on relatively short timescales, and can be strongly
affected by Doppler boosting and beaming. The complex nature of  
outflows from X-ray binaries 
does not allow the direct application of physical principles to 
obtain the beam power of each outflow from detailed studies of that
outflow; a theoretically motivated scaling between radio luminosity and
beam power is used to obtain the beam power. The different source properties
studied may make it difficult to compare results obtained with large extended
radio sources to those obtained with X-ray binaries.

\begin{figure}
    \centering
    \includegraphics[width=80mm]{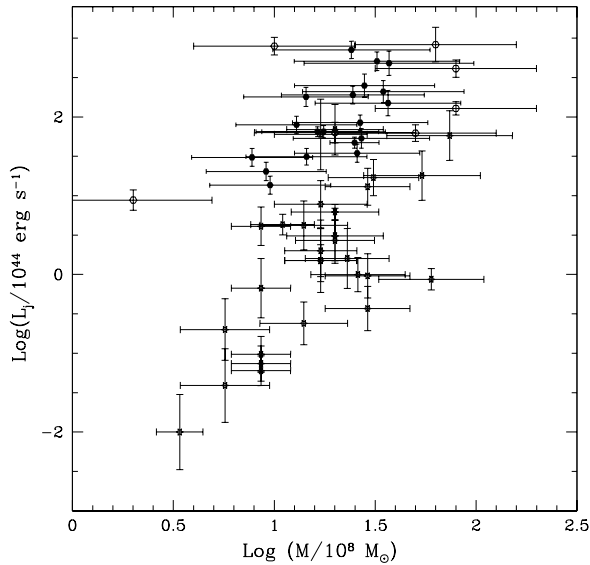}

\caption{Beam powers and black hole masses for 
55 AGN; FRII quasars are shown as open circles, 
FRII galaxies are shown as solid circles, 
and CD sources are shown as open stars.}
	  \label{fig:F1}
    \end{figure}

\subsection{Determination of the Black Hole Spin}

The black hole spin $j$ depends upon the beam power, black hole mass, and magnetic field strength, and is determined for each of the three field strengths considered using equation (1) in the context of the Meier (1999, 2001) model, 
labeled $j_M$.  
Values of $j_M$ and their uncertainties are listed in 
Tables 1 and 2, and are shown in Figures 2 - 7.  
To obtain the fractional 
uncertainty of $j$, the fractional 
uncertainties of the input parameters are
combined in quadrature after noting the following. 
When $B = B_{EDD}$, we have that $j \propto \sqrt{(L_{44}/M_8)}$ (so the 
fractional uncertainty of $j$ is rather small).  
When $B$ is constant, the fractional uncertainty of $j$ is larger 
since $j \propto \sqrt{L_{44}} M_8^{-1}$. When $B\propto j$, 
$j \propto (L_{44})^{1/4} M_8^{-1/2}$, so the fractional uncertainty of $j$ 
is quite small. 

For all three magnetic field strengths,  the black hole spin increases with 
increasing redshift. Fits are obtained using the FRII galaxies and quasars 
only. The FRII sources are from the complete 3CR survey 
(Bennett 1962) including sources from the sample subsets of Laing, Riley, 
\& Longair (1983) and  Pooley et al. (1987); the selection effects for the 
sample are well understood. As mentioned above, the FRII sources studied 
are the most powerful sources at their respective redshifts and define the
envelope of the distribution of sources. 

The slope of $\rm{Log} (j_M)$ as a function of $\rm{Log} (1+z)$ is 
$1.1 \pm 0.2$ for $B=B_{EDD}$;  $0.86 \pm 0.36$  for $B = 10^4$ G;
and $0.43 \pm 0.18$ for $B \propto j$, and are shown in Figures 2-4. 
These slopes are independent of the value of $\kappa$ and thus are valid 
for a broad range of models. 
In all cases, the black hole spin is increasing with redshift. 
The reduced $\chi^2$s of the fits are less than one suggesting that the fits may be a bit more significant than indicated above. 
The quasars and radio galaxies have a similar range of black hole spin at a given redshift, and similar trend with redshift. For the field strengths considered, the FRII sources have spins that range from about 0.2 to 1, and, as noted above, there is a 
consistent trend of spin increasing with increasing redshift. The FRII sources are the most powerful sources at their respective redshifts and thus are likely to define the envelope of the source distribution. The black hole spins 
associated with the CD galaxies studied range from about $10^{-2}$ to about 0.4.  

Figs. 5, 6, and 7 show the black hole spin as a function of black hole mass for the three magnetic field strengths considered.  The FRII galaxies have a small range of black hole mass relative to FRII quasars, and no dependence of spin on black hole mass is evident for the radio galaxies. The FRII quasars have a broader range of black hole mass than the radio galaxies, and the quasar sample is lacking sources with masses that overlap those of the galaxies, so it would be valuable to acquire and study quasars with black hole masses in this range. The CD galaxies have a range of black hole mass that is similar to the FRII galaxy and quasar sample, and a much broader range of spin, with spins extending to very low values. This is likely due to the way that the CD sample is selected.  

Note that in the magnetic switch model of Meier (1999, 2001), outflows are
only produced when the magnetic switch is activated.  Sources with no
outflows can have large spin values; spin values can only be
deduced for sources with outflows.

\begin{figure}
    \centering
    \includegraphics[width=80mm]{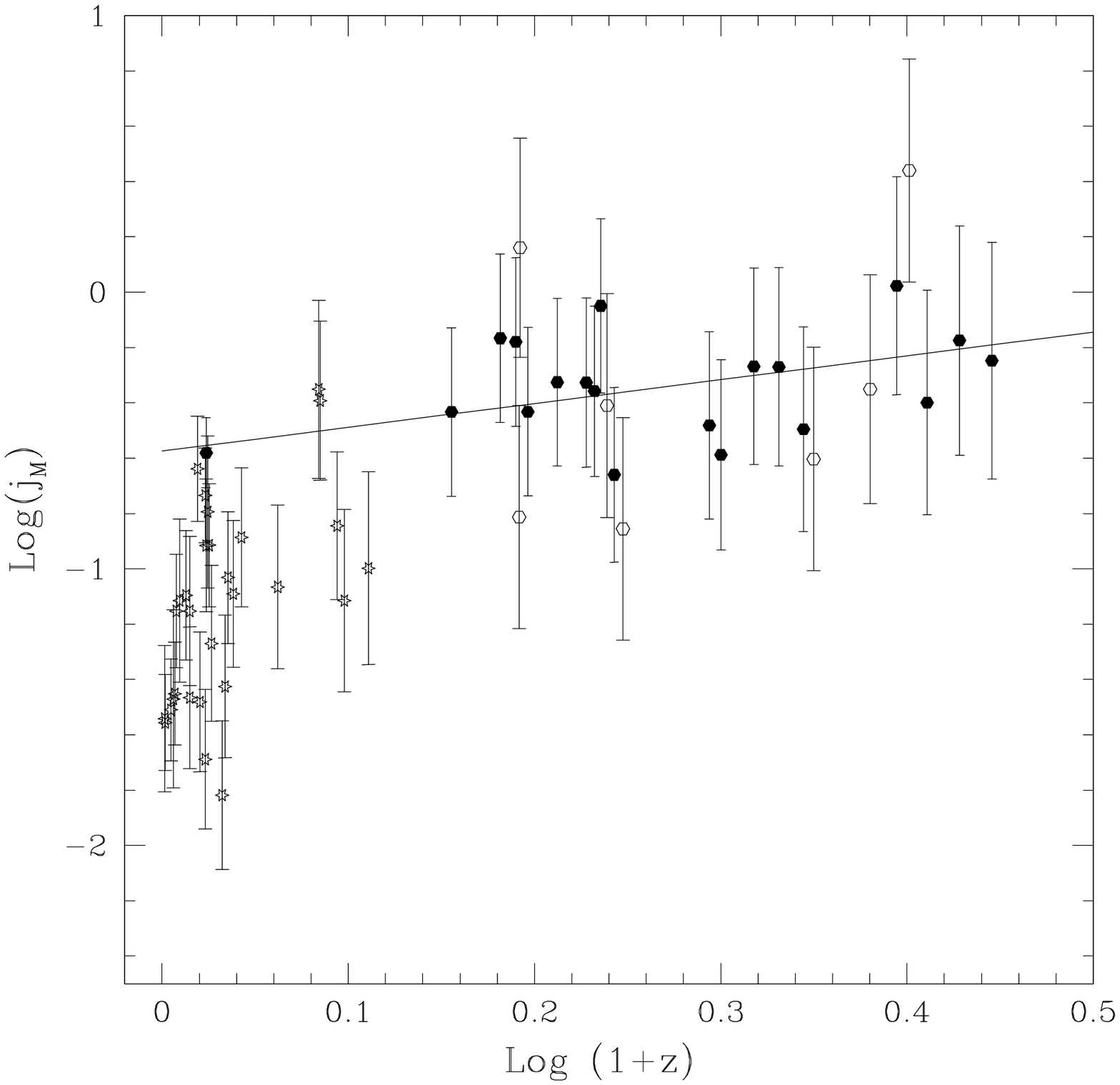}

\caption{Black hole spins obtained with a constant magnetic 
field strength of $10^4$ G; the symbols are as in Fig. \ref{fig:F1}. 
The line is the best fit to the FRII radio galaxies
and quasars only and has a slope of $0.86 \pm 0.36$ and a $\chi^2$ of 13.28 
for 24 degrees of freedom.}
		  \label{fig:F2}
    \end{figure}

\begin{figure}
    \centering
    \includegraphics[width=80mm]{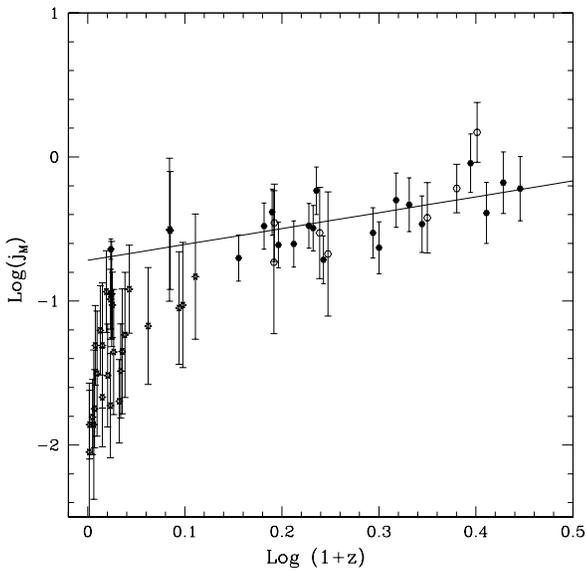}

\caption{Black hole spins obtained with an Eddington magnetic field strength; 
the symbols are as in Fig. \ref{fig:F1}. The line is best fit to the 
FRII radio galaxies
and quasars and has a slope of $1.1 \pm 0.2$ and a $\chi^2$ of 18.0 for 24
degrees of freedom. 
}
		  
\label{fig:F3}
    \end{figure}

\begin{figure}
    \centering
    \includegraphics[width=80mm]{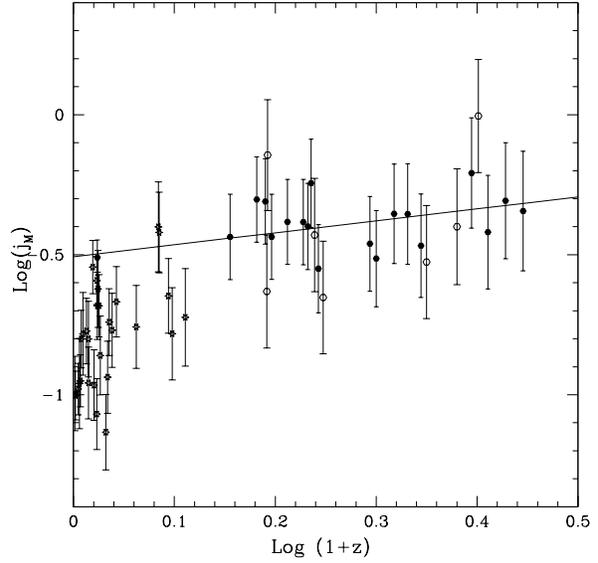}
			 
\caption{Black hole spin obtained with a magnetic field 
strength that is proportional to the spin; 
the symbols are as in Fig. \ref{fig:F1}. 
The line is best fit to the FRII radio galaxies
and quasars and has a slope of $0.43 \pm 0.18$ and a $\chi^2$ 
of 13.2 for 24 degrees of freedom. }
		  \label{fig:F4}
    \end{figure}

\begin{figure}
    \centering
    \includegraphics[width=80mm]{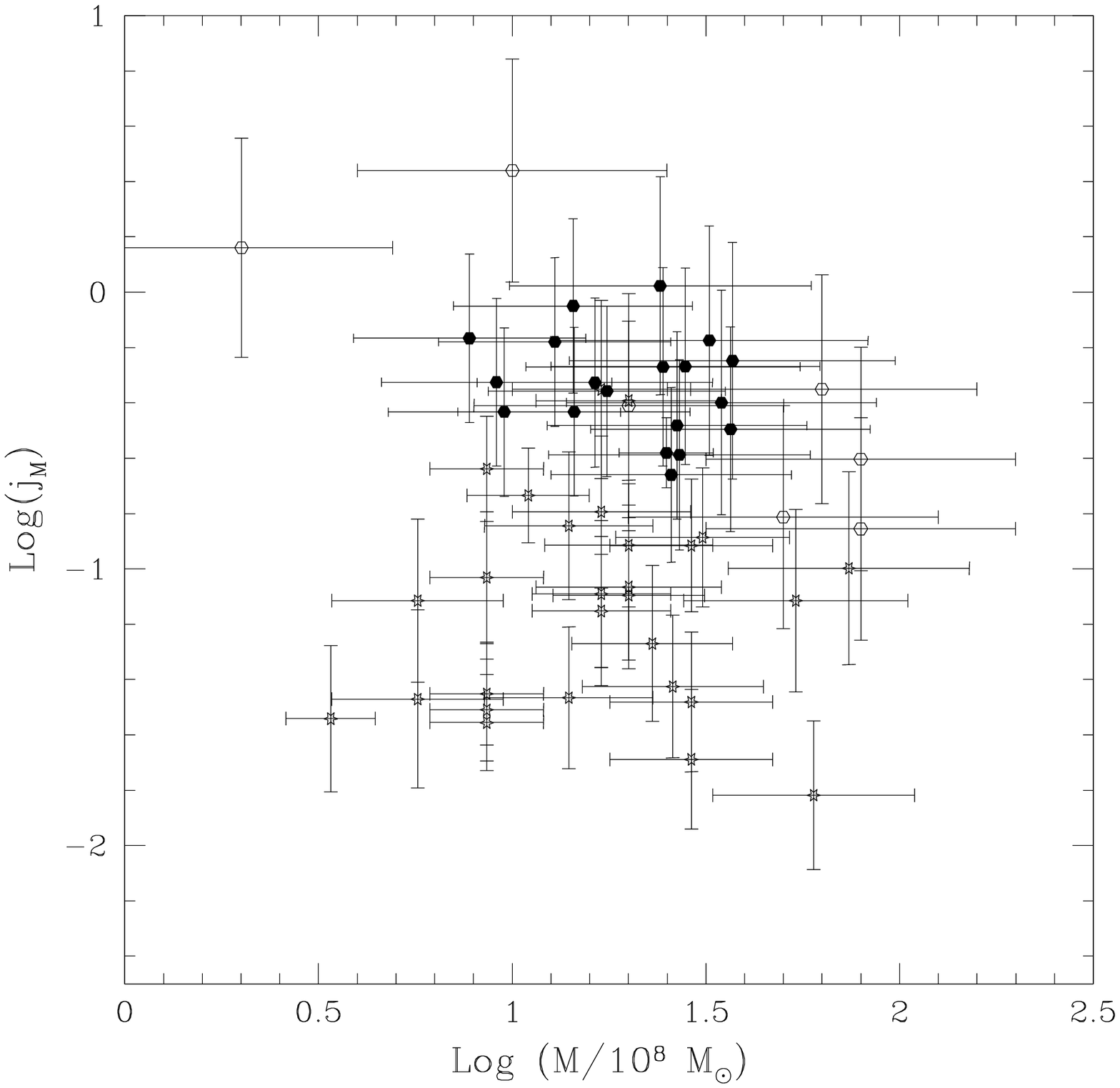}
\caption{As in Fig. \ref{fig:F2} but for spin as a function of black hole mass.}

		  \label{fig:F5}
    \end{figure}

\begin{figure}
    \centering
    \includegraphics[width=80mm]{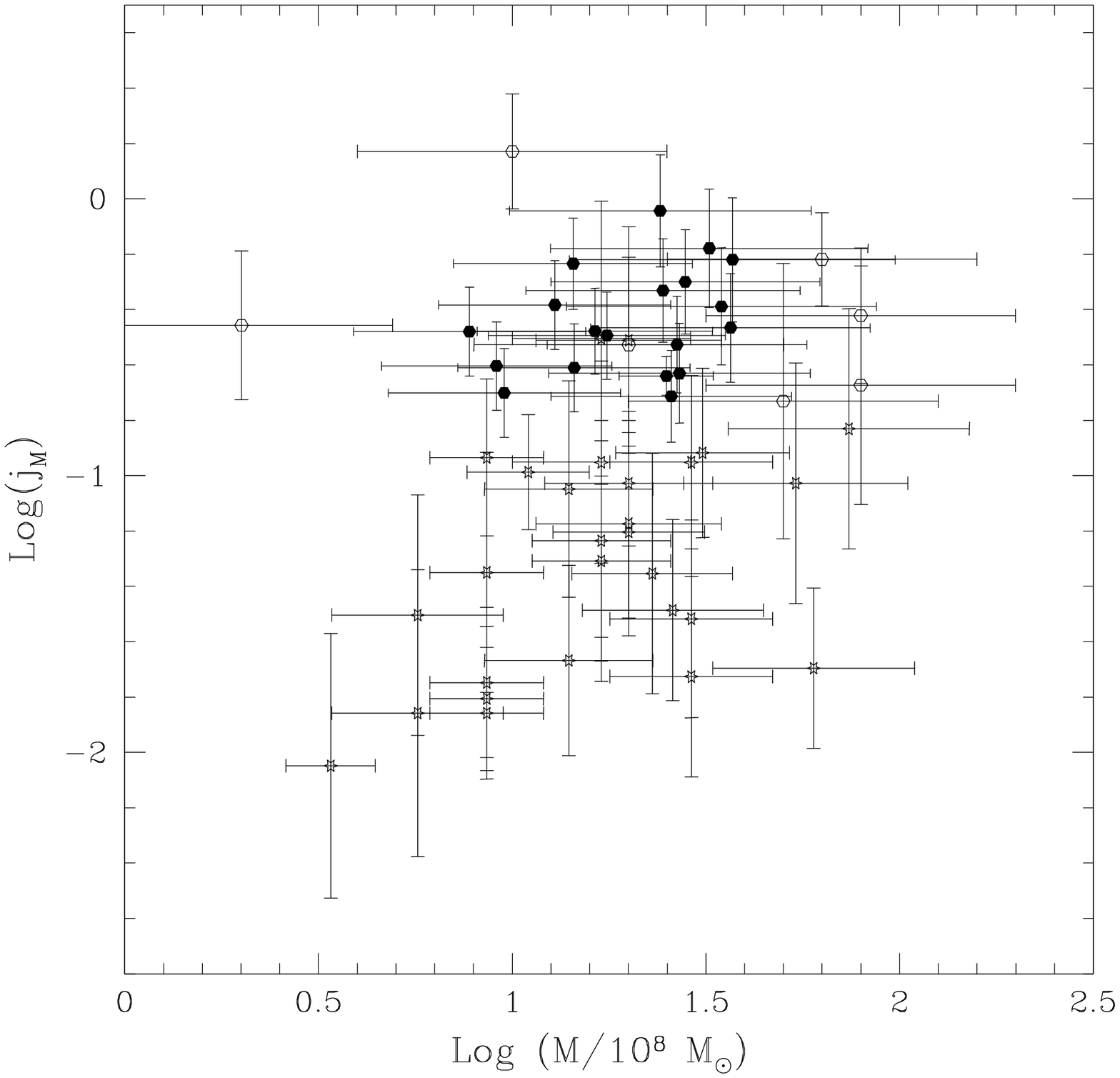}
\caption{As in Fig. \ref{fig:F3} but for spin as a function of black hole mass.}
		 \label{fig:F6}
    \end{figure}

\begin{figure}
    \centering
    \includegraphics[width=80mm]{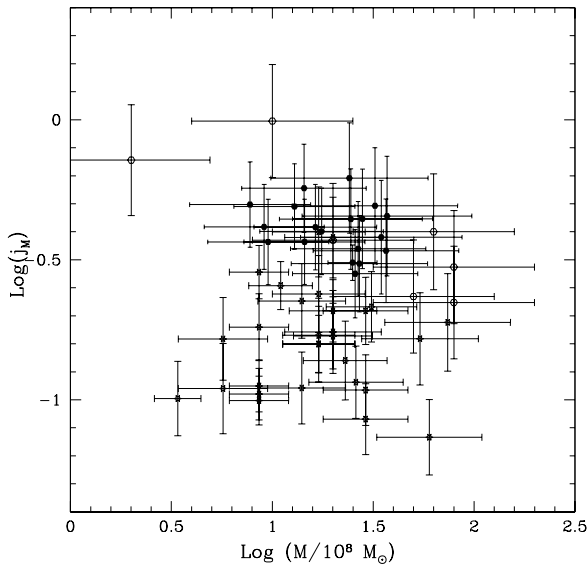}
\caption{As in Fig. \ref{fig:F4} but for spin as a 
function of black hole mass.}
		  \label{fig:F7}
    \end{figure}

\subsection{Determination of the Spin Energy and the Fraction of Spin Energy Extracted}

After the spin $j$ has been determined, the spin energy $E_s$ can be
obtained using equation (2), and the fraction of the spin energy extracted
during a particular outflow event can be obtained using equation (3).  
The spin energy $E_s$ depends upon the spin $j$ and the black hole mass $M$,
while the fraction of the spin energy extracted $f$ depends on these quantities
and the total outflow energy $E_*$. When $j$ is greater than unity,
$E_s$ and $f$ can not be determined (e.g. 3C437, 3C270.1 and 3C275.1).
The values of $E_s/(Mc^2)$, $f$, as well as $E_*/(Mc^2)$ are 
are listed in Tables 3 and 4, and $f$ as a function of redshift
is shown in Figs. 8 and 9. 
Only two magnetic field strengths,  $B=B_{EDD}$ and  $B_4 = 1$, 
are considered here; the conclusion that  $B \propto j$ 
was obtained by assuming that $E_* \propto E_s$ (or $f$ is constant),
as described in section 2,  
so $f$ is, by construction, constant in that case.  
 
The values of $E_s/(Mc^2)$ obtained are all much less than 1. 
This occurs because $r \equiv E_s/(Mc^2)$ is a very steep function of $j$,
ranging from a value of  $r \approx 10^{-3}$ for $j=0.1$ to a 
maximum value of $r \approx 0.29$  
for $j=1$ (e.g. Rees 1984; Blandford 1990), so $r$
is generally small. 
Equation (2) can be re-arranged to write the black hole spin $j$ in
terms of the spin energy per unit black hole mass $r \equiv E_s/(Mc^2)$:
$j = 2(2r-5r^2+4r^3-r^4)^{1/2}$.  
Note that    
by replacing the spin energy $E_s$ with the outflow energy
$E_*$, this expression may used to obtain a lower bound on the black hole
spin that depends only upon the outflow energy and black hole mass
(Daly 2009a). 

The fraction of the spin energy extracted per outflow event
$f$ has a smaller range for the FRII sources studied than for the
CD galaxies studied.  The range of $f$ for FRII sources is about 
0.04 to 0.5 for $B=B_{EDD}$ and 0.03 to 0.3 for $B=10^4$ G,
and FRII quasars have values very similar to FRII galaxies.  
The range of $f$ for CD galaxies is about
0.02 to greater than 1 for $B=B_{EDD}$ and 
0.002 to 0.9 for $B=10^4$ G. 
The values of $f$ for the CD galaxies may be uncertain since 
this quantity depends primarily upon the buoyancy timescale 
for CD galaxies and, 
as discussed by  Rafferty et al. (2006), this is likely to be an
upper limit (see also the discussion in section 3.1).

The redshift evolution of $f$ is studied using the FRII sources only, and
is illustrated in Figs. 8 and 9. For $B=10^4$ G, the best fit slope of 
$\rm{Log}(f)$ as a function of $\rm{Log}(1+z)$ is $-0.52 \pm 0.41$ with a
$\chi^2$ of 15 for 21 degrees of freedom. This is consistent with no evolution 
of $f$ with $z$.  The values of $f$ and redshift evolution 
of FRII quasars is consistent with that of 
FRII galaxies, though the number of quasars is small and 
more data will be needed to confirm this. 
For $B = B_{EDD}$, the best fit slope is $-1.8 \pm 0.4$ with 
a $\chi^2$ of 276 for 23 degrees of freedom. The $\chi^2$ is quite
large; to bring the reduced $\chi^2$ to unity would require increasing the
uncertainty per point by a factor of about 3.5, which would change the
uncertainty of the best fit slope to $-1.8 \pm 1.3$, consistent with no
evolution.  Thus, the data are consistent with no redshift evolution of $f$ 
with redshift with a slight hint that $f$ may decrease with increasing
redshift.  As mentioned earlier, 
$B \propto j$ was obtained by assuming that $f$ is constant 
(Daly \& Guerra 2002; Daly et al. 2009).

\begin{figure}
    \centering
    \includegraphics[width=80mm]{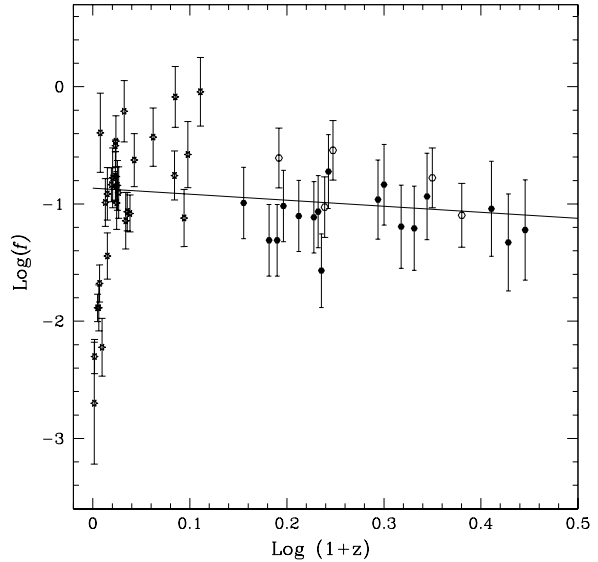}
			 
\caption{Fraction of the spin energy extracted as a function of 
redshift for $B = 10^4$ G;  the symbols are as in Fig. \ref{fig:F1}. 
The best fit line to the FRII radio galaxies 
and quasars is shown and has a slope of 
$-0.52 \pm 0.41$ and a $\chi^2$ of 14.8 for 21
degrees of freedom. This is consistent with no evolution of $f(z)$. }
		  \label{fig:F8}
    \end{figure}

\begin{figure}
    \centering
    \includegraphics[width=80mm]{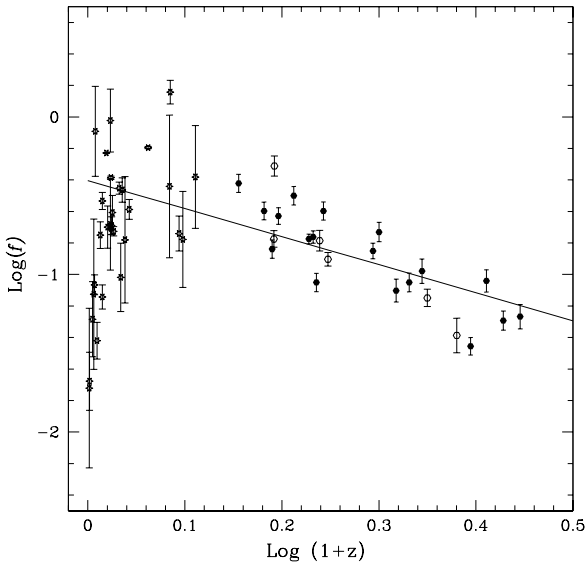}
			 
\caption{Fraction of the spin energy extracted as a function of 
redshift for $B = B_{EDD}$;  the symbols are as in Fig. \ref{fig:F1}. 
The best fit line to the FRII radio galaxies 
and quasars is shown and has a slope of 
$-1.78 \pm 0.37$ and a $\chi^2$ of 276 for 23
degrees of freedom. To bring the reduced $\chi^2$ to unity 
would require increasing the uncertainty per point by about 
3.5, which would change the uncertainty of the best fit slope
to $-1.8 \pm 1.3$, so these data do not require that 
$f(z)$ be strongly evolving with redshift. }
		  \label{fig:F9}
    \end{figure}

\section{Summary and Conclusion}

Black hole spins are estimated for extended radio 
sources fed by outflows from AGN assuming that the beam power of the 
outflow is related to the spin energy of the black hole and surrounding
region. 
Samples of 7 powerful radio loud quasars, 19 powerful radio galaxies,
and 29 CD galaxies with redshifts between about zero and two 
are studied. All of the sources are likely to be  
associated with massive elliptical galaxies. The black hole spins
obtained for powerful FRII quasars and galaxies 
range from about 0.2 to 1 and 
are largest at high redshift and systematically decrease to lower redshift. 
This is the case for each of the three characterizations of 
the magnetic field strength considered.  
The powerful radio galaxies and quasars are drawn from the 3CR sample
and are the most powerful sources at their respective redshifts; 
results obtained with these sources are likely to represent the 
envelope of the distribution. 
The spins of the radio sources associated with CD galaxies range
from about 0.01 to 0.4.  The CD sources are a heterogeneous sample 
and include many nearby low luminosity radio sources.

The normalization of the black hole spin, $\kappa$, varies
by factors of a few for different specific models of spin 
energy extraction from the hole, ergosphere, and surrounding region.  
The trend
of black hole spin increasing with redshift indicated by 
the analysis presented here is independent of the value of $\kappa$; 
different values of $\kappa$ will modify the overall normalization 
of the spins values, changing all of the spins by the same factor. 
If independent spin determinations of one
or more of these sources are obtained, the value of $\kappa$ could
be empirically constrained or determined. 

The decrease of the black hole spin with decreasing redshift 
is consistent with the 
predictions of the models of Hughes \& Blandford (2003) and 
King, Pringle, \& Hofmann (2008).  
This suggests that beam power is directly related to 
black hole spin.  Given the 
indications of the evolution of spin with redshift for 
massive elliptical galaxies obtained here, detailed models 
for the accretion and merger history of the sources and the 
type of accretion that occurs can be constructed, as discussed
by Berti \& Volonteri (2008). 
Clearly, the results presented here are along the lines of the
predictions of King, Pringle, \& Hofmann (2008) in that the
mean spin decreases monotonically with decreasing redshift, 
source to source spins have 
fluctuations of order $0.2$, and the spin tends to 
decrease with increasing black hole mass for the homogeneous 
sample of 3CR sources (though this is rather 
tentative because the black hole mass is an input to the 
spin determination). Consistency between model predictions and
the estimated spins suggests that the beam power is directly
related to black hole spin. 

The fraction of the spin energy extracted per outflow event 
ranges from about 0.03 to 0.5 for FRII 
quasars and galaxies, and from about 0.002 to about one for radio 
sources associated with CD galaxies. The results 
are consistent with this fraction being independent of redshift.  
When this fraction is high, the outflow can significantly modify
the spin of the hole, while when this fraction is low, the outflow
has a small effect on the black hole spin. In this case, the merger
and accretion history of AGN will have a larger impact on black
hole spin evolution than the outflow history.  

The results obtained here may be considered in light of 
the work of Elvis, Risaliti, \& Zamorani (2002), who 
find that black
holes associated with AGN that produce the X-ray background must be
rapidly spinning, on average, at the time that background is produced.  
Models developed by Gilli, Salvati, \& Hasinger (2001) suggest
that the X-ray background is produced by sources with a range
of redshift with much of the background produced
by sources with redshift between 1 and 3. 
The elliptical galaxies studied here are found to have 
high spins at redshift of about 2, 
with spins slowly decreasing to lower redshift.
If the sources that produce the X-ray background have spins that evolve
in a manner similar to those studied here, the results of Elvis,
Risaliti, \& Zamorani (2002) might suggest that the most mild evolution 
of spin with redshift is favored; in the context of the current study,
this would favor the magnetic field strength $B \propto j$, which 
yields spins that evolve as $(1+z)^{0.43 \pm 0.18}$.  Note that this
field strength is also favored by a comparison of the empirical results
of Allen et al. (2006) and Merloni \& Heinz (2007) with 
spin energy extraction models, as described at the end of section 2. 
Another factor that would affect the production of the X-ray background
is the normalization of the spin, through the factor $\kappa$.  If 
the sources that produce the X-ray background have spins similar to 
those studied here, it would suggest a value of $\kappa \simeq \sqrt{5}$ 
as indicated by the BZ model.  In this case, the spins of FRII 
sources vary from values of order unity at a redshift of about 2 to 
about 0.7 at a redshift of about 0 (e.g. Daly 2009b). Given the 
range of redshift of sources contributing to the X-ray background,
and the indications from this study that there is a range of spin 
values at a given redshift for radio sources, a detailed study 
would have to be carried out to determine whether the sources
that produce the X-ray background could have spin normalizations
and redshift evolution similar to those obtained here for radio sources. 

The spin values obtained here for AGN are in good agreement with typical 
spin values reported for X-ray binary systems, which are summarized by 
Fender et al. (2010). Studies of X-ray binaries 
have not indicated a link between black hole 
spin and the properties of the jets  
(Steiner et al. 2010; Fender et al. 2010; Gou et al. 2010), 
except for a correlation between the transient jet normalization 
and spin (Fender et al. 2010).  
According to  Fender et al. (2010), the 
lack of a relationship between 
jet properties and spin for X-ray binaries  
could be due to a problem with the jet properties or speed determinations; a problem 
with the spin measurements; or a real lack of dependence of jet properties
on spin.  
Studies of AGN suggest a link between the beam power of large-scale outflows and black 
hole spin (Sikora, Stawarz, \& Lasota 2007).  
To date, the samples
studied are rather small, and very different parts of the systems  
are used to study the jet properties of X-ray binaries and the beam powers of AGN, 
as described in section 3.1. Thus, it is not clear how significant
the differences are between the relationship of beam power and spin for AGN and 
the relationship of jet properties and spin for 
X-ray binaries. If differences truly exist, it might suggest that 
more than one mechanism may lead to the formation of jets from the 
vicinity of a black hole. In related work,  Fernandes et al. (2010) studied 
powerful radio galaxies, concluding that the 
accreted energy is channeled into jets with maximum efficiency for the most
powerful sources, and suggesting that black hole spin plays a dominant role 
in the production of jets.  
Punsly (2010) compared jet efficiencies with currently popular ``no net flux'' 
models of spin energy extraction 
and found that these models could not account for the empirically determined jet efficiencies. 
The detailed study of Kovacs, Gergely, \& Biermann (2010) suggests that 
efficient jets can be produced in the context of the BZ model of spin energy extraction. 

Overall, the results obtained with powerful 3CR quasars are consistent 
with those obtained with 3CR galaxies.  The sample of quasars studied
is small and larger samples of quasars are needed to quantify this result.
The quasar sample studied has a black hole mass range that is larger than 
that of the 3CR galaxies studied and a dearth of black hole masses that
overlap those of the galaxies. There is a hint that the quasars may
lie at the edges of the distribution of spins for 3CR sources at a 
given redshift, and this might be related to black hole mass.  
This can be tested
with larger quasars samples, and with samples of powerful radio
galaxies with a broader range of black hole mass. 

\section*{Acknowledgments}
It is a pleasure to thank Laura Brenneman, Kris Beckwith,  Preeti Kharb, 
Brian McNamara, Chris Reynolds, and Chris O'Dea for interesting and helpful 
discussions of this work. I would also like to thank the referee
and editor for helpful comments and suggestions.


\begin{table*}
\begin{minipage}{140mm}
\caption{Black Hole Spins of FRII Sources}   
\label{tab:comp}        
\begin{tabular}{llrrrrrrrrr}   
\hline\hline                    
Source &  type& z & {$L_{44}$%
\footnote{Total beam powers are the weighted sum of the 
beam power from each of the two lobes of each source. Input values for 
each side of each radio galaxy (RG) are obtained from by O'Dea et al. (2009), 
who used the data sets 
of Leahy, Muxlow, \& Stephens (1989), Liu, Pooley, \& Riley (1992), 
Guerra, Daly, \& Wan (2000), and new observations. The beam power is
obtained by applying the equations of strong shock physics to the
source. Input values for 
radio loud quasars (RLQ) are obtained from the compilations of Wellman, Daly, 
\& Wan (1997) and Wan, Daly, \& Guerra (2000) after converting quantities
to the cosmological model adopted here; the data used for the study 
originated in the observations of Leahy, Muxlow, \& Stephens (1989) and 
Liu, Pooley, \& Riley (1992). Only sources with black hole mass estimates
are included. }}     
  & {$M_8$%
\footnote{The value for 3C 405 (Cygnus A) is obtained from 
Tadhunter et al. (2003); values for the next four sources are obtained
from McLure et al. (2004); values for the remaining 14 RG are
obtained from McLure et al. (2006). Values for the RLQ are obtained from 
McLure et al. (2006).}}
& $B_{EDD,4}$ &
$j_M(B=B_{EDD})$ & $j_M(B=10^4G)$&$j_M(B \propto j)$
 \\
(1)    &    (2)         &    (3)          &   (4)           &  (5) 
      &       (6)    &(7)&(8)&(9)         \\
\hline                          
                                  3C 405  	&  	RG  	&  	0.056  	&$ 
47  	\pm 	8 	$&$ 	25 	\pm 	7 	$&$ 	1.2 	\pm 	0.2 	$&$ 	0.23 	\pm 
0.04 	$&$ 	0.27 	\pm 	0.08 	$&$ 	0.31 	\pm 	0.05 	$\\
3C 244.1 	& 	RG 	& 	0.43 	&$ 	14 	\pm 	4 	$&$ 	9.5 	\pm 	6.6 	$&$ 
1.9 	\pm 	0.7 	$&$ 	0.20 	\pm 	0.07 	$&$ 	0.38 	\pm 	0.26 	$&$ 	0.37 
\pm 	0.13 	$\\
3C 172 	& 	RG 	& 	0.519 	&$ 	31 	\pm 	8 	$&$ 	7.8 	\pm 	5.4 	$&$ 
2.2 	\pm 	0.7 	$&$ 	0.33 	\pm 	0.12 	$&$ 	0.70 	\pm 	0.49 	$&$ 	0.50 
\pm 	0.17 	$\\
3C 330 	& 	RG 	& 	0.549 	&$ 	80 	\pm 	20 	$&$ 	13 	\pm 	9 	$&$ 	1.7 
\pm 	0.6 	$&$ 	0.41 	\pm 	0.15 	$&$ 	0.68 	\pm 	0.47 	$&$ 	0.49 	\pm 
0.17 	$\\
3C 427.1 	& 	RG 	& 	0.572 	&$ 	31 	\pm 	8 	$&$ 	14 	\pm 	10 	$&$ 
1.6 	\pm 	0.5 	$&$ 	0.25 	\pm 	0.09 	$&$ 	0.38 	\pm 	0.26 	$&$ 	0.37 
\pm 	0.13 	$\\
3C 337 	& 	RG 	& 	0.63 	&$ 	20 	\pm 	6 	$&$ 	9.1 	\pm 	6.2 	$&$ 	2.0 
\pm 	0.7 	$&$ 	0.25 	\pm 	0.09 	$&$ 	0.48 	\pm 	0.34 	$&$ 	0.41 	\pm 
0.14 	$\\
3C34 	& 	RG 	& 	0.69 	&$ 	65 	\pm 	9 	$&$ 	16 	\pm 	11 	$&$ 	1.5 
\pm 	0.5 	$&$ 	0.33 	\pm 	0.12 	$&$ 	0.48 	\pm 	0.34 	$&$ 	0.41 	\pm 
0.15 	$\\
3C441 	& 	RG 	& 	0.707 	&$ 	65 	\pm 	12 	$&$ 	18 	\pm 	12 	$&$ 	1.4 
\pm 	0.5 	$&$ 	0.32 	\pm 	0.12 	$&$ 	0.45 	\pm 	0.32 	$&$ 	0.40 	\pm 
0.14 	$\\
3C 55 	& 	RG 	& 	0.72 	&$ 	180 	\pm 	50 	$&$ 	14 	\pm 	10 	$&$ 	1.6 
\pm 	0.6 	$&$ 	0.58 	\pm 	0.22 	$&$ 	0.91 	\pm 	0.66 	$&$ 	0.57 	\pm 
0.21 	$\\
3C 247 	& 	RG 	& 	0.749 	&$ 	35 	\pm 	9 	$&$ 	26 	\pm 	18 	$&$ 	1.2 
\pm 	0.4 	$&$ 	0.19 	\pm 	0.07 	$&$ 	0.22 	\pm 	0.16 	$&$ 	0.28 	\pm 
0.10 	$\\
3C 289 	& 	RG 	& 	0.967 	&$ 	85 	\pm 	19 	$&$ 	27 	\pm 	21 	$&$ 	1.2 
\pm 	0.4 	$&$ 	0.30 	\pm 	0.12 	$&$ 	0.34 	\pm 	0.26 	$&$ 	0.35 	\pm 
0.13 	$\\
3C 280 	& 	RG 	& 	0.996 	&$ 	53 	\pm 	15 	$&$ 	27 	\pm 	21 	$&$ 	1.2 
\pm 	0.4 	$&$ 	0.23 	\pm 	0.1 	$&$ 	0.26 	\pm 	0.21 	$&$ 	0.31 	\pm 
0.12 	$\\
3C 356 	& 	RG 	& 	1.079 	&$ 	250 	\pm 	85 	$&$ 	28 	\pm 	22 	$&$ 
1.1 	\pm 	0.5 	$&$ 	0.50 	\pm 	0.22 	$&$ 	0.55 	\pm 	0.45 	$&$ 	0.44 
\pm 	0.18 	$\\
3C 267 	& 	RG 	& 	1.144 	&$ 	190 	\pm 	50 	$&$ 	24 	\pm 	20 	$&$ 
1.2 	\pm 	0.5 	$&$ 	0.47 	\pm 	0.2 	$&$ 	0.55 	\pm 	0.45 	$&$ 	0.44 
\pm 	0.18 	$\\
3C 324 	& 	RG 	& 	1.21 	&$ 	150 	\pm 	55 	$&$ 	37 	\pm 	30 	$&$ 	1.0 
\pm 	0.4 	$&$ 	0.34 	\pm 	0.16 	$&$ 	0.33 	\pm 	0.28 	$&$ 	0.34 	\pm 
0.15 	$\\
3C 437 	& 	RG 	& 	1.48 	&$ 	710 	\pm 	180 	$&$ 	24 	\pm 	22 	$&$ 
1.2 	\pm 	0.5 	$&$ 	0.91 	\pm 	0.42 	$&$ 	1.1 	\pm 	1.0 	$&$ 	0.62 
\pm 	0.28 	$\\
3C 68.2 	& 	RG 	& 	1.575 	&$ 	210 	\pm 	70 	$&$ 	35 	\pm 	32 	$&$ 	1.0 
\pm 	0.5 	$&$ 	0.41 	\pm 	0.20 	$&$ 	0.41 	\pm 	0.38 	$&$ 	0.38 	\pm 
0.18 	$\\
3C 322 	& 	RG 	& 	1.681 	&$ 	510 	\pm 	140 	$&$ 	32 	\pm 	30 	$&$ 
1.1 	\pm 	0.5 	$&$ 	0.66 	\pm 	0.33 	$&$ 	0.68 	\pm 	0.65 	$&$ 	0.49 
\pm 	0.24 	$\\
3C 239 	& 	RG 	& 	1.79 	&$ 	480 	\pm 	170 	$&$ 	37 	\pm 	36 	$&$ 	1.0 
\pm 	0.5 	$&$ 	0.60 	\pm 	0.31 	$&$ 	0.58 	\pm 	0.57 	$&$ 	0.45 	\pm 
0.22 	$\\
3C 334 	& 	RLQ 	& 	0.555 	&$ 	62 	\pm 	15 	$&$ 	50 	\pm 	46 	$&$ 
0.8 	\pm 	0.4 	$&$ 	0.19 	\pm 	0.09 	$&$ 	0.15 	\pm 	0.14 	$&$ 	0.23 
\pm 	0.11 	$\\
3C 275.1 	& 	RLQ 	& 	0.557 	&$ 	8.8 	\pm 	2.6 	$&$ 	2.0 	\pm 	1.8 	$&$ 
4.2 	\pm 	1.9 	$&$ 	0.35 	\pm 	0.17 	$&$ 	1.4 	\pm 	1.3 	$&$ 	0.72 
\pm 	0.33 	$\\
3C 254 	& 	RLQ 	& 	0.734 	&$ 	63 	\pm 	20 	$&$ 	20 	\pm 	19 	$&$ 
1.3 	\pm 	0.6 	$&$ 	0.30 	\pm 	0.15 	$&$ 	0.39 	\pm 	0.37 	$&$ 	0.37 
\pm 	0.18 	$\\
3C 175 	& 	RLQ 	& 	0.768 	&$ 	130 	\pm 	30 	$&$ 	79 	\pm 	73 	$&$ 
0.67 	\pm 	0.31 	$&$ 	0.21 	\pm 	0.1 	$&$ 	0.14 	\pm 	0.13 	$&$ 
0.22 	\pm 	0.10 	$\\
3C 68.1 	& 	RLQ 	& 	1.238 	&$ 	410 	\pm 	100 	$&$ 	79 	\pm 	73 	$&$ 
0.68 	\pm 	0.31 	$&$ 	0.38 	\pm 	0.18 	$&$ 	0.25 	\pm 	0.23 	$&$ 
0.30 	\pm 	0.14 	$\\
3C 268.4 	& 	RLQ 	& 	1.4 	&$ 	830 	\pm 	420 	$&$ 	63 	\pm 	58 	$&$ 
0.76 	\pm 	0.35 	$&$ 	0.60 	\pm 	0.32 	$&$ 	0.45 	\pm 	0.43 	$&$ 	0.40 
\pm 	0.19 	$\\
3C 270.1 	& 	RLQ 	& 	1.519 	&$ 	790 	\pm 	210 	$&$ 	10 	\pm 	9 
$&$ 	1.9 	\pm 	0.9 	$&$ 	1.48 	\pm 	0.71 	$&$ 	2.7 	\pm 	2.5 	$&$ 
0.99 	\pm 	0.46 	$\\

\end{tabular}
\end{minipage}
\end{table*}

\begin{table*}
\begin{minipage}{140mm}
\caption{Black Hole Spins of Sources in Galaxy Clusters}   
\label{tab:comp}        
\begin{tabular}{llrrrrrrrrr}   
\hline\hline                    
Source &  {type%
\footnote{Almost all of the sources have FRI or amorphous
radio structure except for a few exceptions such as Cygnus A 
(Birzan et al. 2008).}}
& z & {$L_{44}$%
\footnote{Beam powers are obtained from Rafferty et al. (2006)
who combine the total cavity pressure and volume with the buoyancy timescale
to determine the beam power. }}
  & {$M_8$%
\footnote{Black hole masses are obtained from the values  
$M_{BH,L_K}$ listed in Table 3 of Rafferty et al. (2006) after removing
the 0.35 correction factor introduced in 
that paper.  This brings the black hole mass estimates of Cygnus A 
(3C 405) and M84 into reasonably good 
agreement with the independent determinations of 
Tadhunter et al. (2003) and Maciejewski \& Binney (2001), respectively. 
}}
& $B_{EDD,4}$ &
$j_M(B=B_{EDD})$ & $j_M(B=10^4G)$&$j_M(B \propto j)$
 \\
(1)    &    (2)         &    (3)          &   (4)           &  (5) 
      &       (6)      &(7)&(8)&(9)       \\
\hline

        M84          	& 	CD 	& 	0.0035 	&$ 	0.01 	\pm 	0.01 	$&$ 
3.4 	\pm 	0.9 	$&$ 	3.2 	\pm 	0.4 	$&$ 	0.009 	\pm 	0.005 	$&$ 
0.028 	\pm 	0.017 	$&$ 	0.10 	\pm 	0.03 	$\\
        M87          	& 	CD 	& 	0.0042 	&$ 	0.06 	\pm 	0.03 	$&$ 
8.6 	\pm 	2.9 	$&$ 	2.0 	\pm 	0.3 	$&$ 	0.014 	\pm 	0.004 	$&$ 	0.028 
\pm 	0.011 	$&$ 	0.10 	\pm 	0.02 	$\\
        Centaurus    	& 	CD 	& 	0.011 	&$ 	0.074 	\pm 	0.04 	$&$ 
8.6 	\pm 	2.9 	$&$ 	2.0 	\pm 	0.3 	$&$ 	0.015 	\pm 	0.005 	$&$ 	0.031 
\pm 	0.013 	$&$ 	0.10 	\pm 	0.02 	$\\
        HCG 62       	& 	CD 	& 	0.014 	&$ 	0.039 	\pm 	0.04 	$&$ 
5.7 	\pm 	2.9 	$&$ 	2.5 	\pm 	0.6 	$&$ 	0.014 	\pm 	0.008 	$&$ 
0.034 	\pm 	0.025 	$&$ 	0.11 	\pm 	0.04 	$\\
        A262         	& 	CD 	& 	0.016 	&$ 	0.097 	\pm 	0.05 	$&$ 
8.6 	\pm 	2.9 	$&$ 	2.0 	\pm 	0.3 	$&$ 	0.018 	\pm 	0.005 	$&$ 	0.035 
\pm 	0.015 	$&$ 	0.11 	\pm 	0.02 	$\\
        Perseus      	& 	CD 	& 	0.018 	&$ 	1.5 	\pm 	0.7 	$&$ 	17 
\pm 	7 	$&$ 	1.4 	\pm 	0.3 	$&$ 	0.049 	\pm 	0.015 	$&$ 	0.070 	\pm 
0.033 	$&$ 	0.16 	\pm 	0.04 	$\\
        PKS 1404-267 	& 	CD 	& 	0.022 	&$ 	0.20 	\pm 	0.18 	$&$ 	5.7 
\pm 	2.9 	$&$ 	2.5 	\pm 	0.6 	$&$ 	0.031 	\pm 	0.016 	$&$ 	0.076 	\pm 
0.051 	$&$ 	0.16 	\pm 	0.05 	$\\
        A2199        	& 	CD 	& 	0.03 	&$ 	2.7 	\pm 	1.6 	$&$ 	20 
\pm 	9 	$&$ 	1.3 	\pm 	0.3 	$&$ 	0.061 	\pm 	0.022 	$&$ 	0.080 	\pm 
0.041 	$&$ 	0.17 	\pm 	0.04 	$\\
        A2052        	& 	CD 	& 	0.035 	&$ 	1.5 	\pm 	1.4 	$&$ 	17 
\pm 	7 	$&$ 	1.4 	\pm 	0.3 	$&$ 	0.049 	\pm 	0.024 	$&$ 	0.070 	\pm 
0.043 	$&$ 	0.16 	\pm 	0.05 	$\\
        2A 0335+096  	& 	CD 	& 	0.035 	&$ 	0.24 	\pm 	0.15 	$&$ 
14 	\pm 	7 	$&$ 	1.6 	\pm 	0.4 	$&$ 	0.022 	\pm 	0.008 	$&$ 	0.033 
\pm 	0.020 	$&$ 	0.11 	\pm 	0.03 	$\\
        MKW 3S       	& 	CD 	& 	0.045 	&$ 	4.1 	\pm 	2.3 	$&$ 	8.6 
\pm 	2.9 	$&$ 	2.0 	\pm 	0.3 	$&$ 	0.12 	\pm 	0.04 	$&$ 	0.23 	\pm 
0.10 	$&$ 	0.29 	\pm 	0.06 	$\\
        A4059        	& 	CD 	& 	0.048 	&$ 	0.96 	\pm 	0.62 	$&$ 	29 
\pm 	14 	$&$ 	1.1 	\pm 	0.3 	$&$ 	0.031 	\pm 	0.012 	$&$ 	0.033 	\pm 
0.020 	$&$ 	0.11 	\pm 	0.03 	$\\
        Hydra A      	& 	CD 	& 	0.055 	&$ 	4.3 	\pm 	1.3 	$&$ 	11 
\pm 	4 	$&$ 	1.8 	\pm 	0.3 	$&$ 	0.10 	\pm 	0.02 	$&$ 	0.18 	\pm 
0.07 	$&$ 	0.25 	\pm 	0.05 	$\\
        A85          	& 	CD 	& 	0.055 	&$ 	0.37 	\pm 	0.24 	$&$ 	29 
\pm 	14 	$&$ 	1.1 	\pm 	0.3 	$&$ 	0.019 	\pm 	0.008 	$&$ 	0.021 	\pm 
0.012 	$&$ 	0.086 	\pm 	0.026 	$\\
        Cygnus A     	& 	CD 	& 	0.056 	&$ 	13 	\pm 	7 	$&$ 	29 	\pm 
14 	$&$ 	1.1 	\pm 	0.3 	$&$ 	0.11 	\pm 	0.04 	$&$ 	0.12 	\pm 	0.07 
$&$ 	0.21 	\pm 	0.06 	$\\
        Sersic 159/03 	& 	CD 	& 	0.058 	&$ 	7.8 	\pm 	5.4 	$&$ 	17 
\pm 	9 	$&$ 	1.4 	\pm 	0.4 	$&$ 	0.11 	\pm 	0.05 	$&$ 	0.16 	\pm 
0.10 	$&$ 	0.24 	\pm 	0.07 	$\\
        A133         	& 	CD 	& 	0.06 	&$ 	6.2 	\pm 	1.4 	$&$ 	20 
\pm 	10 	$&$ 	1.3 	\pm 	0.3 	$&$ 	0.093 	\pm 	0.025 	$&$ 	0.12 	\pm 
0.06 	$&$ 	0.21 	\pm 	0.05 	$\\
        A1795        	& 	CD 	& 	0.063 	&$ 	1.6 	\pm 	1.4 	$&$ 	23 
\pm 	11 	$&$ 	1.3 	\pm 	0.3 	$&$ 	0.044 	\pm 	0.022 	$&$ 	0.054 	\pm 
0.036 	$&$ 	0.14 	\pm 	0.05 	$\\
        A2029        	& 	CD 	& 	0.077 	&$ 	0.87 	\pm 	0.27 	$&$ 	60 
\pm 	36 	$&$ 	0.77 	\pm 	0.23 	$&$ 	0.020 	\pm 	0.007 	$&$ 	0.015 	\pm 
0.009 	$&$ 	0.073 	\pm 	0.023 	$\\
        A478         	& 	CD 	& 	0.081 	&$ 	1.0 	\pm 	0.5 	$&$ 	26 	\pm 
14 	$&$ 	1.2 	\pm 	0.3 	$&$ 	0.033 	\pm 	0.012 	$&$ 	0.038 	\pm 
0.023 	$&$ 	0.12 	\pm 	0.04 	$\\
        A2597        	& 	CD 	& 	0.085 	&$ 	0.67 	\pm 	0.58 	$&$ 	8.6 
\pm 	2.9 	$&$ 	2.0 	\pm 	0.3 	$&$ 	0.047 	\pm 	0.022 	$&$ 	0.093 	\pm 
0.051 	$&$ 	0.18 	\pm 	0.05 	$\\
        3C 388       	& 	CD 	& 	0.092 	&$ 	2.0 	\pm 	1.8 	$&$ 	17 	\pm 
7 	$&$ 	1.4 	\pm 	0.3 	$&$ 	0.057 	\pm 	0.028 	$&$ 	0.081 	\pm 
0.049 	$&$ 	0.17 	\pm 	0.05 	$\\
        PKS 0745-191 	& 	CD 	& 	0.103 	&$ 	17 	\pm 	9 	$&$ 	31 	\pm 
16 	$&$ 	1.1 	\pm 	0.3 	$&$ 	0.12 	\pm 	0.04 	$&$ 	0.13 	\pm 	0.07 
$&$ 	0.21 	\pm 	0.06 	$\\
        Hercules A   	& 	CD 	& 	0.154 	&$ 	3.1 	\pm 	2.5 	$&$ 	20 
\pm 	11 	$&$ 	1.3 	\pm 	0.4 	$&$ 	0.066 	\pm 	0.032 	$&$ 	0.086 	\pm 
0.060 	$&$ 	0.17 	\pm 	0.06 	$\\
        Zw 2701      	& 	CD 	& 	0.214 	&$ 	60 	\pm 	62 	$&$ 	17 	\pm 
9 	$&$ 	1.4 	\pm 	0.4 	$&$ 	0.31 	\pm 	0.18 	$&$ 	0.44 	\pm 	0.32 
$&$ 	0.40 	\pm 	0.14 	$\\
        MS 0735.6+7421 	& 	CD 	& 	0.216 	&$ 	69 	\pm 	51 	$&$ 	20 
\pm 	11 	$&$ 	1.3 	\pm 	0.4 	$&$ 	0.31 	\pm 	0.14 	$&$ 	0.41 	\pm 
0.28 	$&$ 	0.38 	\pm 	0.13 	$\\
        4C 55.16     	& 	CD 	& 	0.242 	&$ 	4.2 	\pm 	3.0 	$&$ 	14 	\pm 
7 	$&$ 	1.6 	\pm 	0.4 	$&$ 	0.09 	\pm 	0.039 	$&$ 	0.14 	\pm 	0.09 
$&$ 	0.22 	\pm 	0.07 	$\\
        A1835        	& 	CD 	& 	0.253 	&$ 	18 	\pm 	13 	$&$ 	54 	\pm 
36 	$&$ 	0.81 	\pm 	0.27 	$&$ 	0.10 	\pm 	0.05 	$&$ 	0.076 	\pm 
0.057 	$&$ 	0.16 	\pm 	0.06 	$\\
        Zw 3146      	& 	CD 	& 	0.291 	&$ 	58 	\pm 	42 	$&$ 	74 	\pm 
53 	$&$ 	0.70 	\pm 	0.25 	$&$ 	0.15 	\pm 	0.07 	$&$ 	0.10 	\pm 	0.08 
$&$ 	0.19 	\pm 	0.08 	$\\
\end{tabular}
\end{minipage}
\end{table*}

\begin{table*}
\begin{minipage}{140mm}
\caption{FRII Black Hole Spin Energy and Fraction of Spin Energy Extracted}   
\label{tab:comp}        
\begin{tabular}{llccccc}   
\hline\hline                    
Source &  type&  {${E_* \over Mc^2}(10^{-3})$%
\footnote{The total outflow energy $E_*$ is taken to be the weighted
sum of the outflow energy of each side of a source; 
the values for the RG were obtained from Daly (2009a) who 
used the outflow energies listed by O'Dea et al. 
(2009) for the radio galaxies; values for RLQ were obtained in 
an identical manner. }}  
& ${E_s(B=B_{EDD}) 
\over Mc^2}$ 
& $f(B=B_{EDD})$ & ${E_s(B=10^4 G) \over Mc^2}$ & $f(B=10^4 G)$ \\
(1)    &    (2)         &    (3)          &   (4)           &  (5) 
      &       (6)      &(7)       \\
\hline                          
		
3C 405	&	RG	&$	1.3	\pm	0.4	$&$	0.0067	\pm	0.0022	$&$	0.195	\pm	0.016	$&$
0.0092	\pm	0.0055	$&$	0.141	\pm	0.041	$\\
3C 244.1	&	RG	&$	1.9	\pm	1.4	$&$	0.0050	\pm	0.0037	$&$	0.379	\pm	0.050	$&$
0.019	\pm	0.027	$&$	0.102	\pm	0.072	$\\
3C 172	&	RG	&$	3.6	\pm	2.6	$&$	0.014	\pm	0.011	$&$	0.253	\pm	0.033	$&$
0.07	\pm	0.13	$&$	0.049	\pm	0.034	$\\
3C 330	&	RG	&$	3.3	\pm	2.4	$&$	0.023	\pm	0.018	$&$	0.145	\pm	0.019	$&$
0.07	\pm	0.12	$&$	0.049	\pm	0.034	$\\
3C 427.1	&	RG	&$	1.8	\pm	1.3	$&$	0.0077	\pm	0.0057	$&$	0.235	\pm	0.029	$&$
0.019	\pm	0.027	$&$	0.096	\pm	0.067	$\\
3C 337	&	RG	&$	2.5	\pm	1.8	$&$	0.0079	\pm	0.0059	$&$	0.316	\pm	0.043	$&$
0.032	\pm	0.048	$&$	0.079	\pm	0.055	$\\
3C34	&	RG	&$	2.4	\pm	1.7	$&$	0.014	\pm	0.011	$&$	0.168	\pm	0.012	$&$	0.031
\pm	0.048	$&$	0.077	\pm	0.054	$\\
3C441	&	RG	&$	2.3	\pm	1.7	$&$	0.013	\pm	0.010	$&$	0.173	\pm	0.015	$&$
0.027	\pm	0.041	$&$	0.086	\pm	0.061	$\\
3C 55	&	RG	&$	4.3	\pm	3.2	$&$	0.048	\pm	0.042	$&$	0.089	\pm	0.012	$&$	0.16
\pm	0.44	$&$	0.027	\pm	0.020	$\\
3C 247	&	RG	&$	1.2	\pm	0.9	$&$	0.0047	\pm	0.0037	$&$	0.253	\pm	0.034	$&$
0.0063	\pm	0.0094	$&$	0.19	\pm	0.14	$\\
3C 289	&	RG	&$	1.6	\pm	1.3	$&$	0.0113	\pm	0.0094	$&$	0.141	\pm	0.016	$&$
0.015	\pm	0.024	$&$	0.109	\pm	0.085	$\\
3C 280	&	RG	&$	1.3	\pm	1.1	$&$	0.0070	\pm	0.0059	$&$	0.186	\pm	0.026	$&$
0.009	\pm	0.014	$&$	0.15	\pm	0.12	$\\
3C 356	&	RG	&$	2.7	\pm	2.3	$&$	0.034	\pm	0.033	$&$	0.079	\pm	0.013	$&$
0.042	\pm	0.078	$&$	0.064	\pm	0.052	$\\
3C 267	&	RG	&$	2.6	\pm	2.2	$&$	0.029	\pm	0.027	$&$	0.089	\pm	0.012	$&$
0.042	\pm	0.078	$&$	0.062	\pm	0.051	$\\
3C 324	&	RG	&$	1.6	\pm	1.4	$&$	0.015	\pm	0.014	$&$	0.105	\pm	0.019	$&$
0.014	\pm	0.024	$&$	0.116	\pm	0.099	$\\
3C 437	&	RG	&$	5.5	\pm	5.1	$&$	0.16	\pm	0.27	$&$	0.0350	\pm	0.0045	$& ---& ---\\
3C 68.2	&	RG	&$	2	\pm	1.9	$&$	0.022	\pm	0.023	$&$	0.091	\pm	0.015	$&$
0.022	\pm	0.043	$&$	0.091	\pm	0.085	$\\
3C 322	&	RG	&$	3.3	\pm	3.2	$&$	0.065	\pm	0.077	$&$	0.0510	\pm	0.0072	$&$
0.07	\pm	0.16	$&$	0.047	\pm	0.045	$\\
3C 239	&	RG	&$	2.8	\pm	2.8	$&$	0.052	\pm	0.062	$&$	0.054	\pm	0.010	$&$
0.05	\pm	0.11	$&$	0.060	\pm	0.059	$\\
3C 270.1	&	RLQ	&$	13.1	\pm	7.7	$&	---	&	---	&	---& ---\\
3C 268.4	&	RLQ	&$	2.1	\pm	1.3	$&$	0.052	\pm	0.064	$&$	0.041	\pm	0.010	$&$
0.027	\pm	0.055	$&$	0.080	\pm	0.050	$\\
3C 68.1	&	RLQ	&$	1.33	\pm	0.78	$&$	0.019	\pm	0.019	$&$	0.0710	\pm	0.0090
$&$	0.008	\pm	0.015	$&$	0.167	\pm	0.098	$\\
3C 175	&	RLQ	&$	0.71	\pm	0.41	$&$	0.0057	\pm	0.0055	$&$	0.125	\pm	0.012
$&$	0.0025	\pm	0.0046	$&$	0.29	\pm	0.17	$\\
3C 254	&	RLQ	&$	1.9	\pm	1.1	$&$	0.011	\pm	0.012	$&$	0.164	\pm	0.025	$&$
0.020	\pm	0.040	$&$	0.094	\pm	0.056	$\\
3C 275.1	&	RLQ	&$	7.7	\pm	4.6	$&$	0.016	\pm	0.016	$&$	0.488	\pm	0.071	$& --- & ---\\
3C 334	&	RLQ	&$	0.73	\pm	0.43	$&$	0.0044	\pm	0.0042	$&$	0.168	\pm	0.020
$&$	0.0030	\pm	0.0056	$&$	0.25	\pm	0.15	$\\
																																	
\end{tabular}
\end{minipage}
\end{table*}

\begin{table*}
\begin{minipage}{140mm}
\caption{CD Black Hole Spin Energy and Fraction of Spin Energy Extracted}   
\label{tab:comp}        
\begin{tabular}{lccccc}   
\hline\hline                    
Source &    {${E_* \over Mc^2}(10^{-3})$%
\footnote{The total energy is obtained from Rafferty et al. (2006).}}
& ${E_s(B=B_{EDD}) \over Mc^2}$ 
& $f(B=B_{EDD})$ & ${E_s(B=10^4 G) \over Mc^2}$ & $f(B=10^4 G)$ \\
(1)    &    (2)         &    (3)          &   (4)           &  (5) 
      &       (6)          \\
\hline         
				
        M84         	&$	0.00019	\pm	0.00023	$&$	0.000010	\pm	0.000011	$&$	0.019
\pm	0.022	$&$	0.00010	\pm	0.00012	$&$	0.0020	\pm	0.0024	$\\
        M87         	&$	0.00052	\pm	0.00028	$&$	0.000024	\pm	0.000013
$&$	0.0210	\pm	0.0089	$&$	0.000097	\pm	0.000077	$&$	0.0050	\pm	0.0017	$\\
        Centaurus   	&$	0.0016	\pm	0.0010	$&$	0.000030	\pm	0.000018	$&$	0.052
\pm	0.029	$&$	0.00012	\pm	0.00010	$&$	0.0130	\pm	0.0035	$\\
        HCG 62      	&$	0.0018	\pm	0.0022	$&$	0.000024	\pm	0.000028	$&$	0.075
\pm	0.082	$&$	0.00014	\pm	0.00021	$&$	0.0130	\pm	0.0059	$\\
        A262        	&$	0.0034	\pm	0.0020	$&$	0.000039	\pm	0.000024	$&$	0.086
\pm	0.012	$&$	0.00016	\pm	0.00013	$&$	0.0210	\pm	0.0076	$\\
        Perseus     	&$	0.25	\pm	0.19	$&$	0.00030	\pm	0.00018	$&$	0.81	\pm	0.53
$&$	0.00061	\pm	0.00057	$&$	0.41	\pm	0.32	$\\
        PKS 1404-267	&$	0.0047	\pm	0.0045	$&$	0.00012	\pm	0.00012	$&$	0.038	\pm
0.010	$&$	0.00073	\pm	0.00097	$&$	0.0060	\pm	0.0034	$\\
        A2199       	&$	0.083	\pm	0.057	$&$	0.00047	\pm	0.00034	$&$	0.178	\pm
0.035	$&$	0.00081	\pm	0.00083	$&$	0.103	\pm	0.049	$\\
        A2052       	&$	0.022	\pm	0.022	$&$	0.00030	\pm	0.00030	$&$	0.072	\pm
0.013	$&$	0.00061	\pm	0.00075	$&$	0.036	\pm	0.016	$\\
        2A 0335+096 	&$	0.017	\pm	0.013	$&$	0.000058	\pm	0.000046	$&$	0.293	\pm
0.037	$&$	0.00014	\pm	0.00016	$&$	0.122	\pm	0.063	$\\
        MKW 3S      	&$	0.99	\pm	0.65	$&$	0.0017	\pm	0.0011	$&$	0.591	\pm	0.005
$&$	0.0068	\pm	0.0060	$&$	0.146	\pm	0.049	$\\
        A4059       	&$	0.023	\pm	0.018	$&$	0.00012	\pm	0.00010	$&$	0.200	\pm
0.062	$&$	0.00014	\pm	0.00017	$&$	0.167	\pm	0.098	$\\
        Hydra A     	&$	1.24	\pm	0.74	$&$	0.00131	\pm	0.00062	$&$	0.95	\pm	0.44
$&$	0.0040	\pm	0.0032	$&$	0.314	\pm	0.035	$\\
        A85         	&$	0.0093	\pm	0.0078	$&$	0.000045	\pm	0.000037	$&$	0.21
\pm	0.14	$&$	0.000054	\pm	0.000064	$&$	0.173	\pm	0.082	$\\
        Cygnus A    	&$	0.65	\pm	0.46	$&$	0.0016	\pm	0.0011	$&$	0.412	\pm	0.010
$&$	0.0019	\pm	0.0021	$&$	0.34	\pm	0.17	$\\
        Sersic 159/03	&$	0.32	\pm	0.27	$&$	0.0016	\pm	0.0014	$&$	0.204	\pm
0.027	$&$	0.0032	\pm	0.0039	$&$	0.102	\pm	0.053	$\\
        A133        	&$	0.27	\pm	0.15	$&$	0.00108	\pm	0.00059	$&$	0.247	\pm
0.062	$&$	0.0019	\pm	0.0019	$&$	0.144	\pm	0.070	$\\
        A1795       	&$	0.046	\pm	0.046	$&$	0.00024	\pm	0.00025	$&$	0.188	\pm
0.013	$&$	0.00037	\pm	0.00049	$&$	0.125	\pm	0.063	$\\
        A2029       	&$	0.018	\pm	0.012	$&$	0.000050	\pm	0.000034	$&$	0.353	\pm
0.031	$&$	0.000029	\pm	0.000035	$&$	0.62	\pm	0.37	$\\
        A478        	&$	0.013	\pm	0.010	$&$	0.00014	\pm	0.00010	$&$	0.096	\pm
0.048	$&$	0.00018	\pm	0.00022	$&$	0.072	\pm	0.040	$\\
        A2597       	&$	0.093	\pm	0.085	$&$	0.00027	\pm	0.00025	$&$	0.344	\pm
0.061	$&$	0.0011	\pm	0.0012	$&$	0.086	\pm	0.033	$\\
        3C 388      	&$	0.067	\pm	0.068	$&$	0.00041	\pm	0.00040	$&$	0.17	\pm
0.15	$&$	0.0008	\pm	0.0010	$&$	0.083	\pm	0.030	$\\
        PKS 0745-191	&$	0.49	\pm	0.34	$&$	0.0019	\pm	0.0013	$&$	0.259	\pm	0.038
$&$	0.0021	\pm	0.0023	$&$	0.24	\pm	0.12	$\\
        Hercules A  	&$	0.34	\pm	0.34	$&$	0.00054	\pm	0.00053	$&$	0.639	\pm
0.010	$&$	0.0009	\pm	0.0013	$&$	0.37	\pm	0.21	$\\
        Zw 2701     	&$	4.5	\pm	5.2	$&$	0.013	\pm	0.015	$&$	0.36	\pm	0.38	$&$
0.026	\pm	0.040	$&$	0.175	\pm	0.084	$\\
        MS 0735.6+7421	&$	18	\pm	16	$&$	0.012	\pm	0.012	$&$	1.44	\pm	0.25	$&$
0.022	\pm	0.031	$&$	0.82	\pm	0.49	$\\
        4C 55.16    	&$	0.19	\pm	0.16	$&$	0.00102	\pm	0.00089	$&$	0.182	\pm
0.047	$&$	0.0025	\pm	0.0030	$&$	0.076	\pm	0.043	$\\
        A1835       	&$	0.19	\pm	0.19	$&$	0.0012	\pm	0.0011	$&$	0.17	\pm	0.12
$&$	0.0007	\pm	0.0011	$&$	0.26	\pm	0.17	$\\
        Zw 3146     	&$	1.1	\pm	1.2	$&$	0.0027	\pm	0.0028	$&$	0.42	\pm	0.31	$&$
0.0013	\pm	0.0020	$&$	0.91	\pm	0.61	$\\

\end{tabular}
\end{minipage}
\end{table*}
\end{document}